\documentclass[twocolumn]{openjournal}
\usepackage{graphicx}
\usepackage{amsmath}
\usepackage{amssymb}
\usepackage{xspace}
\usepackage{color}
\usepackage[hidelinks]{hyperref}
\usepackage{orcidlink}

\newcommand{\y}[1]{y^{(#1)}}

\begin{document}

\title{A new timestep criterion for N-body simulations}

\shorttitle{A new timestep criterion for N-body simulations}
\shortauthors{Pham, Rein \& Spiegel}

\author{\vspace{-1.2cm}Dang Pham\,\orcidlink{0000-0002-0924-8403}}
\affiliation{Department of Astronomy and Astrophysics, University of Toronto, Toronto, Ontario, M5S 3H4, Canada}

\author{Hanno Rein$^{*}$\,\orcidlink{0000-0003-1927-731X}}
\affiliation{Department of Physical and Environmental Sciences, University of Toronto at Scarborough, Toronto, Ontario M1C 1A4, Canada}
\affiliation{Department of Astronomy and Astrophysics, University of Toronto, Toronto, Ontario, M5S 3H4, Canada}
\email{${}^{*}$\text{hanno.rein@utoronto.ca}}

\author{David S. Spiegel\,\orcidlink{0000-0002-1830-4104}}
\affiliation{Google\\}

\begin{abstract}
We derive a new criterion for estimating characteristic dynamical timescales in N-body simulations.
The criterion uses the second, third, and fourth derivatives of particle positions: acceleration, jerk, and snap.
It can be used for choosing timesteps in integrators with adaptive step size control.
For any two-body problem the criterion is guaranteed to determine the orbital period and pericenter timescale regardless of eccentricity.
We discuss why our criterion is the simplest derivative-based expression for choosing adaptive timesteps with the above properties and show its superior performance over existing criteria in numerical tests.
Because our criterion uses lower order derivatives, it is less susceptible to rounding errors caused by finite floating point precision.
This significantly decreases the volume of phase space where an adaptive integrator fails or gets stuck due to unphysical timestep estimates.
For example, our new criterion can accurately estimate timesteps for orbits around a 50m sized Solar System object located at 40AU from the coordinate origin when using double floating point precision.
Previous methods where limited to objects larger than 10km.
We implement our new criterion in the high order IAS15 integrator which is part of the freely available N-body package REBOUND.
\end{abstract}


\maketitle



\section{Introduction}
In numerical integrations of the N-body problem, one looks for approximate solutions to a set of coupled second-order differential equations for some initial conditions.
In this process, one typically discretizes time into finite timesteps.
In many cases, such as the evolution of a planetary system in which the planet's orbits don't change significantly, a fixed timestep is sufficient.
In fact, many algorithms, especially symplectic ones such as Wisdom-Holman type splitting schemes \citep{WisdomHolman1991} require a fixed timestep for good performance over long timescales.
If so, then {\it a priori} knowledge about the solution and the desired accuracy can guide the choice of timestep. 
Typically, the timestep is a small fraction of the shortest dynamical timescale in the problem, i.e. the orbital period or the pericenter timescale \citep{Wisdom2015}.

However, there are also many cases where a fixed timestep is not adequate.
Consider once again a planetary system, but this time one that becomes unstable after some time so that planets have a close encounter with each other.
In such a case, the shortest dynamical timescale suddenly becomes very small during the close encounter.
Using a fixed timestep would require the timestep to be prohibitively small for the entire duration of the simulation, just to resolve one encounter.
Adaptive methods can solve this problem by automatically adjusting the timestep such that all physically relevant timescales are always resolved.

How to choose the timestep in practice is the topic of this paper. 
We focus our discussion on IAS15, a 15th-order integrator widely used for gravitational dynamics \citep{ReinSpiegel2015}. 
IAS15 is based on a Gauß-Radau quadrature \citep{Everhart1985} and comes with an adaptive step size control that can automatically choose a timestep.
IAS15 is part of the REBOUND package \citep{ReinLiu2012} and has enabled hundreds of scientific studies covering a wide range of dynamical systems, in particular those which are not well suited for symplectic integration methods\footnote{Note that IAS15 converges to machine precision in most cases. If so, then IAS15 is in practice no less symplectic than an integrator which is symplectic on paper.}.
However, note that our results are general and the criterion we derive in Sect.~\ref{sec:newscheme} can be used in other integrators as well, for example those aimed at simulations of stellar clusters \citep[e.g.][]{NitadoriAarseth2012,WangSpurzem2015}.

We start our discussion with a short summary of existing timestepping criteria often used in N-body simulations in Sect.~\ref{sec:otherschemes}.
Then, in Sect.~\ref{sec:floatingpoint}, we point out how rounding error due to finite floating point precision arise in the existing criterion used in IAS15.
We derive and explore our new proposed timestepping criterion in Sect.~\ref{sec:newscheme}.
By focusing on a two-body system we analytically show that our new criterion is resolving the physically relevant timescales, especially in the limit of high eccentricity. 
In Sect.~\ref{sec:tests} we run numerical integrations with IAS15 and the new criterion.
Most importantly, we will show that close encounters can now be integrated reliably even if they occur far from the coordinate origin.
We summarize and conclude in Sec.~\ref{sec:conclusions}.

\section{Other timestep criteria}
\label{sec:otherschemes}
We begin with a review of commonly used timestep criteria used in N-body simulations.
We use a notation where $y$ denotes the particle coordinates. 
For readability, we ignore subtleties such as particle and coordinate indices, but come back to this issue at the end of this section.
We denote time derivatives of $y$ as follows
\begin{eqnarray}
    \y{n} \equiv \frac{d^n}{dt^n} y .
\end{eqnarray}
In principle any ratio of derivatives $\y{m}/\y{n}$ can be used to calculate a timescale $\tau$:
\begin{eqnarray}
    \tau_{m,n} \equiv \left(\frac{\y{m}}{\y{n}}\right)^{1/(n-m)}\quad\quad \forall n>m\geq 2. \label{eq:tau}
\end{eqnarray}
Note that $\tau$ always has units of time. 
Once we have a characteristic timescale, we can multiply it by a constant dimensionless number $\eta$ to calculate a timestep $dt= \eta \cdot \tau$ which can then be used in integration methods with adaptive step size control. 
The numerical value of $\eta$ depends on the integration method and the required precision.

The simplest choice to define a timescale is to use the ratio of acceleration over jerk (the derivative of the acceleration):
\begin{eqnarray}
    \tau_{\mathrm{2,3}} \equiv
    \frac{\y{2} }{\y{3}}.
\end{eqnarray}
This formula uses the lowest order derivatives: one cannot use the position or the velocity as it would not preserve Galilean invariance (such a criterion would use a different timestep whenever the system is boosted to a new frame moving at a constant velocity). 
Alternatively, some authors, e.g. \cite{PressSpergel1988}, choose the ratio of acceleration over snap (the second derivative of the acceleration) to define 
\begin{eqnarray}
    \tau_{\mathrm{2,4}} \equiv
    \sqrt{\frac{\y{2} }{\y{4}}}.
\end{eqnarray}
The original IAS15 integrator \citep{ReinSpiegel2015}, the one we aim to improve in this paper, defines a timescale using the 2nd and 9th derivate of the position:
\begin{eqnarray}
    \tau_{\mathrm{RS15}} \equiv \sqrt[7]{\frac{\y{2} }{\y{9}}} \label{eq:olddt}
\end{eqnarray}
with a default value\footnote{The IAS15 implementation uses an accuracy control parameter~$\epsilon$, which is related to $\eta$ by $\epsilon = \eta^7/5400$.} of $\eta = \sqrt[7]{5.04 \cdot 10^{-6}}$.
The reasoning given by \cite{ReinSpiegel2015} for choosing the 9th derivative came from the fact that it is the last term in a series expansion used internally by IAS15 which allowed the authors to estimate the error of the integration scheme.

Hermite integrators \citep{MakinoAarseth1992} and Adams-Bashforth-Moulton schemes \citep{Aarseth1963,Aarseth1985} are other popular tools for N-body simulations.
The timestep criterion used by \cite{Aarseth1985} is 
\begin{eqnarray}
    \tau_{\mathrm{A85}} \equiv \sqrt{ 
    \frac{\y{2} \y{4} + \y{3}\y{3}}{\y{3}\y{5}+\y{4}\y{4}}  
    }.\label{eq:A85}
\end{eqnarray}
The author states that this expression was derived ``after some experimentation'' but other than its good performance in numerical tests, we are unaware of any justification for this specific choice.
Nevertheless, \cite{Aarseth1985} correctly points out several features of this expression: it is independent of mass for two body motion and well defined in special cases such as $\y{2}=0$ or $\y{3}=\y{5}=0$ with all particles at rest.
In Sec.~\ref{sec:newscheme} below, we provide further arguments as to why this expression works well, in particular for dominant two-body motion.

Note that in all the above formula as well as the ones below, we assume that some norm is applied to the $3N$ dimensional derivative vectors $\y{n}$ to obtain a scalar value. 
The precise choice of norm is not crucial for the discussion in this paper.
Often an $L^2$ norm, an $L^\infty$ norm, or a combination of the two is used.
For example, if $y^{(n)}_{a,i}$ refers to the $n$-th derivative of the Cartesian coordinate $i$ and particle with index $a$, then a reasonable choice of norm is

\begin{eqnarray}
  \left| y^{(n)} \right|  = \max_a \sqrt{\sum_{i=x,y,z} \left(y^{(n)}_{a,i}\right)^2}.
\end{eqnarray}

\section{Finite floating point precision}
\label{sec:floatingpoint}
Internally, IAS15 uses a series expansion to approximate trajectories during a timestep.
As mentioned above, \cite{ReinSpiegel2015} used the highest order term in the series expansion to obtain an error estimate for the current step and then propose an optimal timestep for the next step. 
Although this choice was well motivated, when implemented using finite floating point precision it is not numerically stable in many cases.
The reason is that high order derivatives in IAS15 are estimated with some finite difference method\footnote{This is also true for other integration methods.}.
The precise way this is done is not important for this discussion and we refer the reader to the IAS15 paper for the details \citep[see also][]{Everhart1985}.
However, as an illustration consider the following central finite difference scheme for the first derivative:
\begin{eqnarray}
\y{1}(t) \approx \frac{y(t+dt) - y(t-dt)}{2dt} \nonumber 
\end{eqnarray}
The calculation of the difference in the numerator can lead to problems because $dt$ is typically small and thus $y(t+dt)\approx y(t-dt)$. 
Working in double floating point precision where we have about 16 decimal digits of precision to work with, the best accuracy for the estimate of the derivative is obtained by choosing $dt$ such that
\begin{eqnarray}
\left|\frac{y(t+dt) - y(t-dt) } {y(t)}\right| \approx 10^{-8}.
\end{eqnarray}
Choosing a larger $dt$ increases the discretization error from the finite difference scheme. 
Choosing a smaller $dt$ increases the rounding error from finite floating point precision. 
Now consider the case where $y$ is offset by a constant $\bar y$.
This happens frequently in N-body simulations.
For example it is common in a simulation of the Solar System to place the Sun at the origin. 
Then, whenever an asteroid has a close encounter with a planet it will occur at some offset $\bar y$ from the origin.
The close encounter distance can be very small (e.g. the Earth's radius) compared to $\bar y$ (e.g. the Earth's semi-major axis). 
In such a case, we don't have 16 decimal digits of precision anymore. 
In the case of a close encounter with the Earth, we have $\sim 12$ digits left to work with.
As a result estimates of derivatives can quickly become dominated by errors from finite floating point precision. 
Once the estimate for the derivative is no longer accurate, the timestep can become excessively small or excessively large, leading to stalled simulations or unphysical simulation results.

The problem described above gets worse for higher order derivatives and we quantify the effect for IAS15 in Section~\ref{sec:tests}.
This is the main motivation why we are looking to find a criterion which uses the lowest order derivatives possible.

\section{New timestepping criterion}
\label{sec:newscheme}
We now motivate the specific combination of derivatives we choose for our new timestepping criterion.
Let us start by exploring how derivatives behave in a simple eccentric two body system. 
We do this because the motion of a particle in an N-body simulation, at least the kind we are interested in for this paper, is almost always dominated by the interaction with one other particle.
For example to first approximation the Sun orbits the Milky Way, the Earth orbits the Sun, the Moon orbits the Earth, a spacecraft orbits the Moon, and so forth.
There are of course exceptions, but they often don't last long.
Consider, for instance, the motion of a spacecraft when it leaves the Earth's Hill sphere on its way to the Moon: early in its journey, it's effectively in a hyperbolic orbit of Earth; late in its journey, it's effectively in a bound orbit of the Moon; only near the Hill sphere is it neither approximately in orbit of the Earth nor the Moon.
In such a case a particle can at some point experience zero net acceleration.
For as long as a particle is travelling along a straight line it has no physical timescale associated with its trajectory.
In practice, this issue is resolved by making sure the timestep cannot increase by more than some fixed ratio from one step to the next\footnote{This is called a safety factor in IAS15 and we use a numerical value of~0.25 to limit the increase and decrease of the timestep in consecutive steps.}.

To start, let us identify two characteristic timescales in the two body problem: the orbital period $T_p$ and the pericenter timescale $T_f$, which we define as 
\begin{eqnarray}
T_p &\equiv& \frac{2\pi}n \quad\quad\quad \text{ and}\\
T_f &\equiv& \frac{2\pi}n \frac{(1-e)^2}{\sqrt{1-e^2}},    \label{eq:peri}
\end{eqnarray}
where $n$ is the mean motion \citep{Wisdom2015}. For circular orbits, the timescales are identical.
In the limit of $e\to1$, the pericenter timescale follows a power law, $T_f \propto (1-e)^{3/2}$.

\begin{figure}
    \centering
    \includegraphics[width=\columnwidth]{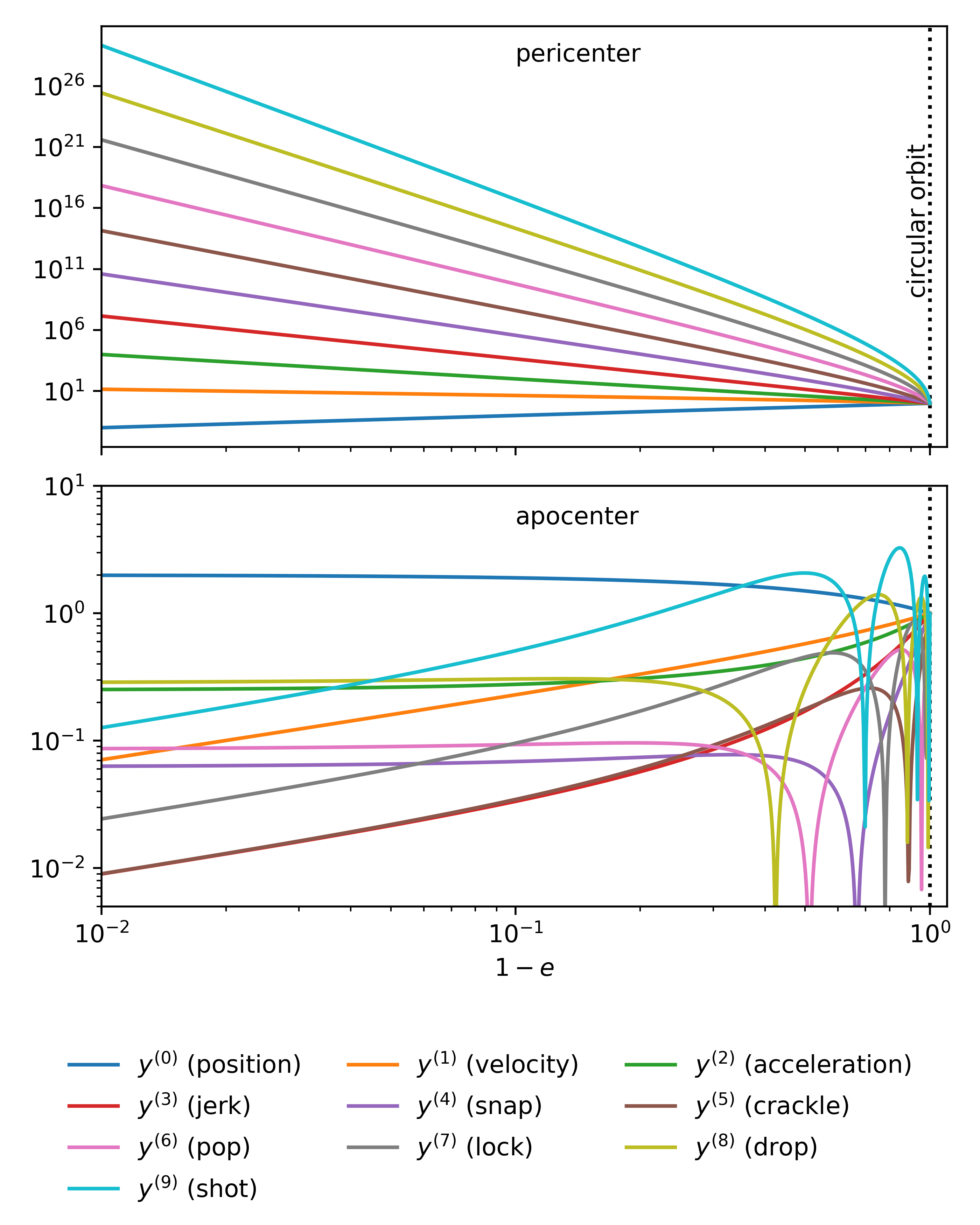}
    \caption{The position $y$ and its first 9 derivatives in the eccentric two body problem. The top panel shows the values at pericenter, the bottom panel at apocenter. We use dimensionless units in which $G=a=1$.}
    \label{fig:derivatives}
\end{figure}

Fig.~\ref{fig:derivatives} shows the absolute values of the position and its derivatives up to $\y{9}$ as a function of the eccentricity at pericenter (top) and apocenter (bottom).
Highly eccentric orbits are on the left-hand side, circular orbits are on the right.
We use dimensionless units such that $G=a=n=1$.
It is easy to show that at pericenter, all derivatives $\y{n}$ follow a power-law in the limit of $e\to1$ with a power-law index of
\begin{eqnarray}
    c_n \equiv \lim_{e \to 1} \frac{\log{\y{n}}   }{\log{(1-e)}} = 1-\frac32 n. \label{eq:cn}
\end{eqnarray}
Similarly, one can show that at apocenter the power-law index in the limit of $e\to1$ is
\begin{eqnarray}
    d_n \equiv \lim_{e \to 1} \frac{\log{\y{n}}   }{\log{(1-e)}} =  
    \begin{cases}
    0 & \text{for } n \text{ even} \\
    \frac12 & \text{for } n \text{ odd} 
  \end{cases}. \label{eq:dn}
\end{eqnarray}
These fundamental results have several important implications for timestepping criteria.

First, note that up to a constant factor, any ratio of successive derivatives in the form of $\tau_{m,m+1}=\y{m}/\y{m+1}$ can be used to calculate the pericenter timescale. 
At pericenter, these ratios are well behaved in both limits $e\to1$ and $e \to 0$. 
The situation is more complicated at apocenter. 
At apocenter, the pericenter timescale is irrelevant from a timestepping point of view. 
However, it is imperative to keep resolving the orbital timescale even if a particle moves very slowly while it is close to apocenter.
We cannot use the ratio of successive derivatives, $\tau_{m,m+1}$ as defined in Eq.~\ref{eq:tau}, to calculate the orbital timescale because in the limit of $e\to1$, $\tau_{m,m+1}$ tends to either~$0$ or~$\infty$ but never to a finite constant.
This is counter to the orbital timescale which is of course independent of $e$. 
What we can do instead, is to use $\tau_{m,m+2}$ which does converge to a constant in the limit of $e\to1$ for all even $m \ge 2$.
Note that we cannot use $m=0$ or~$m=1$ because of Galilean invariance. 
We therefore have to use at least the fourth derivative $\y{4}$ to be able to find a well defined orbital timescale for all eccentricities at apocenter. 
As one can see in Fig.~\ref{fig:derivatives}, this leads to another problem:
Every $\y{n}$ with $n\geq 4$ has at least one root crossing at some $e'$.
For example, although the timescale $\tau_{2,4}$ is well behaved in the limit of $e\to1$, it is ill defined for at least one finite $e'\approx 0.4$.

\begin{figure}
    \centering
    \includegraphics[width=\columnwidth]{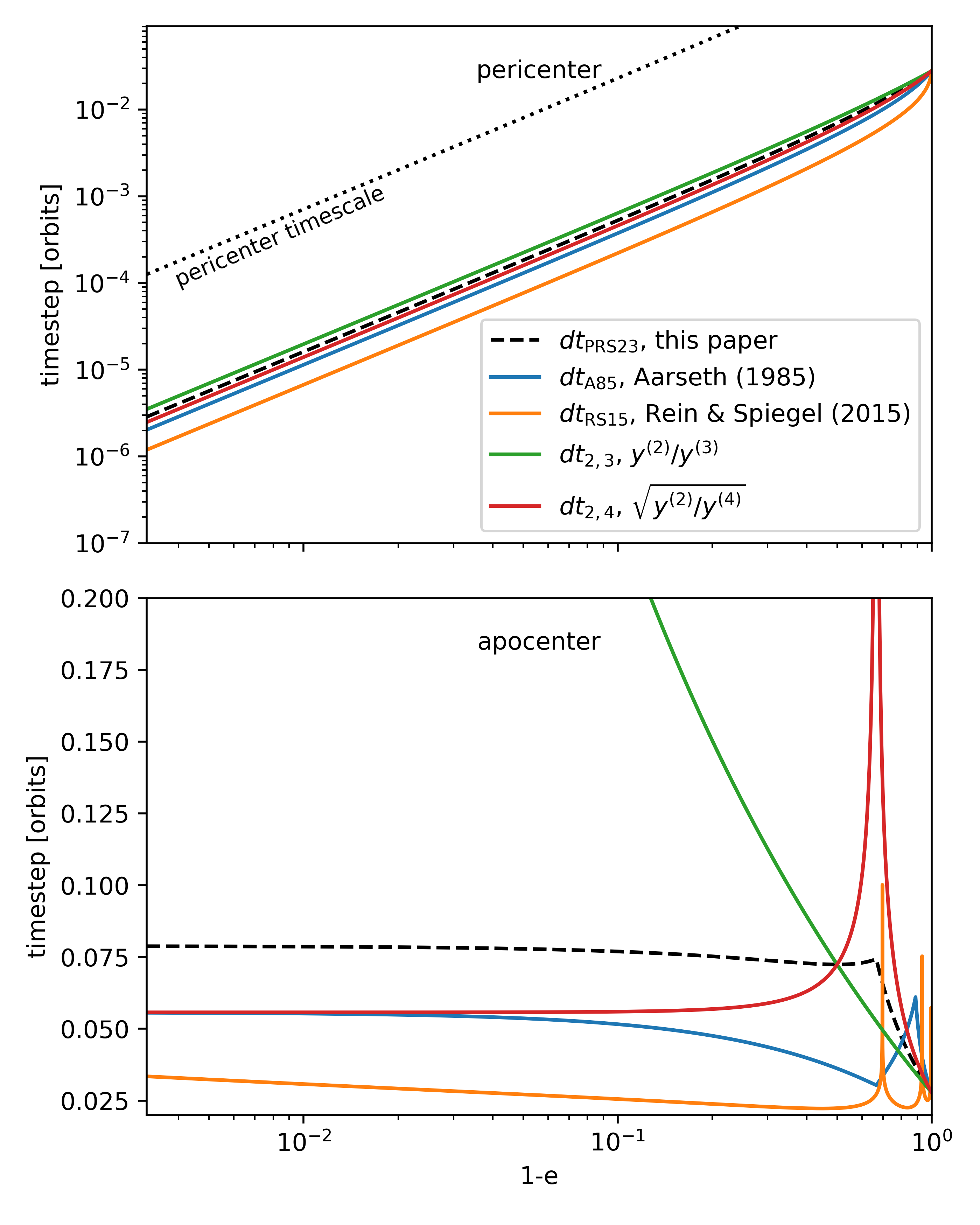}
    \caption{Timesteps as a function of eccentricity in an eccentric two body problem as determined by different timestepping criteria. The top panel shows the timesteps at pericenter, the bottom panel shows the timesteps at apocenter. Also shown is the pericenter timescale, defined such that it is $2\pi$ for circular orbits.}
    \label{fig:timesteps}
\end{figure}

\begin{figure}
    \centering
    \includegraphics[width=\columnwidth]{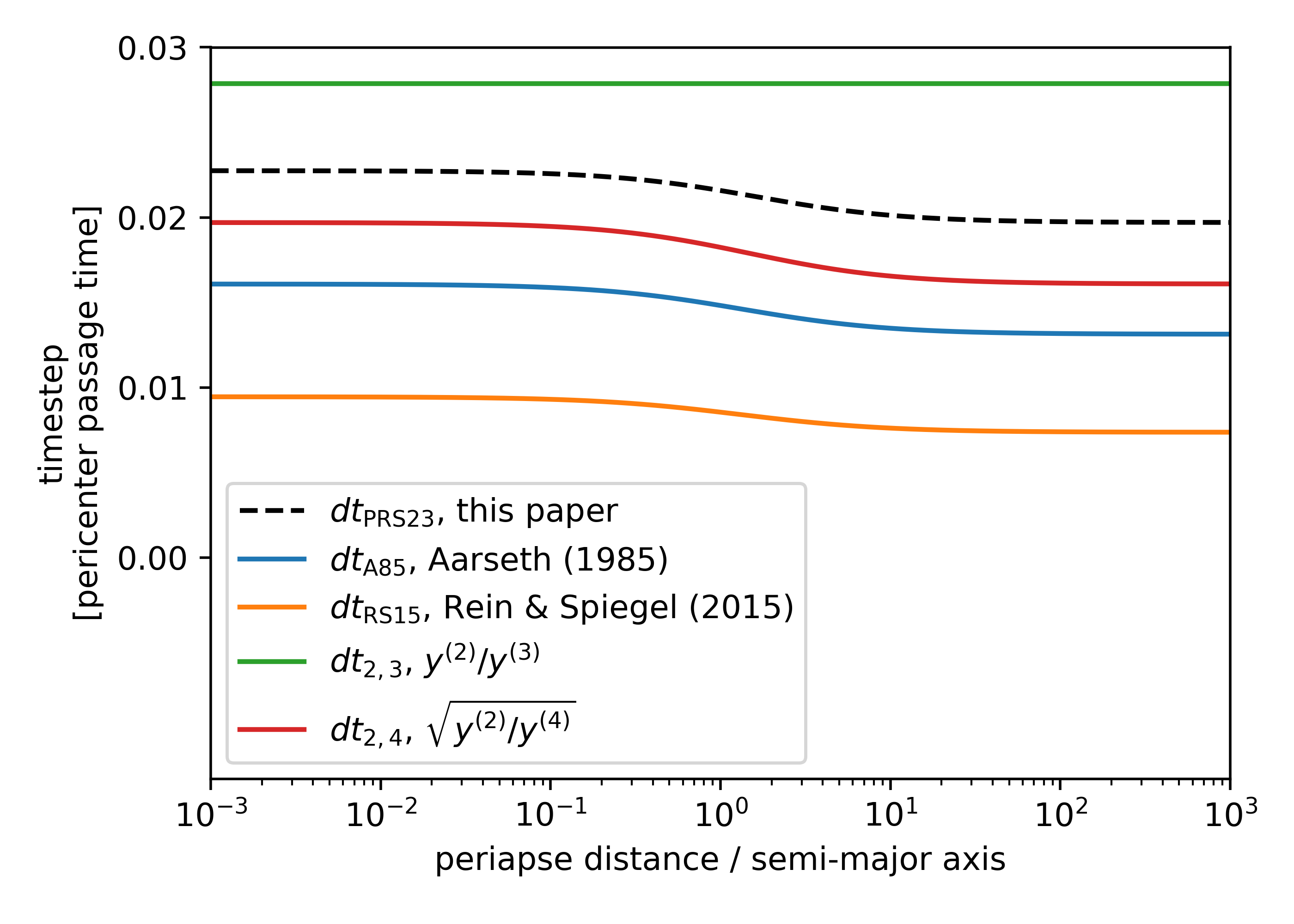}
    \caption{
    Same as Fig.~\ref{fig:timesteps} but for hyperbolic orbits at pericenter. Because there hyperbolic orbits do not have an orbital period, timesteps are shown in units of the pericenter passage timescale.}
    \label{fig:timestepshyper}
\end{figure}

With this in mind, we can now understand further failure modes of the original IAS15 timestepping criterion in addition to the finite floating point precision issue mentioned in Section~\ref{sec:floatingpoint}. 
Recall that the original IAS15 timesstepping criterion uses $\sqrt[7]{\y{9}/\y{2}}$ to estimate a timescale. 
Fig.~\ref{fig:timesteps} shows the timestep using this old timestepping criterion at pericenter (orange line in the top panel) and apocenter (bottom panel). 
First, notice that the pericenter is always well resolved:
\begin{eqnarray}
    \lim_{e\to1} \frac{dt_{\mathrm{RS15}}}{T_f} \approx 0.01 \quad\quad\text{    at pericenter.}
\end{eqnarray}
This is expected given the argument above and Eq.~\ref{eq:cn}.
Second, note the spikes at apocenter for low eccentricity orbits. 
These are due to $\y{9}$ having zero crossings, visible in the bottom panel of Fig.~\ref{fig:derivatives}.
Third, note that the timestep increases at apocenter as the eccentricity gets larger.
The timestep does not converge and arbitrarily large timesteps are in principle possible:
\begin{eqnarray}
    \lim_{e\to1} \frac{dt_{\mathrm{RS15}}}{T_P} = \infty \quad\quad\text{    at apocenter.}
\end{eqnarray}
Once again, we can now understand that this is expected given Eq.~\ref{eq:dn}: $\y{2}$ and $\y{9}$ converge to different power laws for large eccentricities at apocenter.

We also plot the timesteps defined by $\tau_{2,3}$ and $\tau_{2,4}$ in Fig.~\ref{fig:timesteps} (green and red lines).
Both criteria are able to accurately keep track of the pericenter timescale. 
Again, this is expected given Eq.~\ref{eq:cn}.
Notice that the timestep $dt_{2/3}$ diverges at apocenter, making it an unreliable choice for even moderately eccentric orbits.
As we have shown above, the timestep $dt_{2/4}$ does not diverge at apocenter.
However, it diverges at some $e'\approx 0.4$.

We also show $dt_\mathrm{A85}$ in Figure~\ref{fig:timesteps} (blue line).
This criterion solves all of the above issues. 
We are now in a position to provide a justification as to why.
Notice that, just like the other criteria, it estimates the pericenter timescale accurately.
This is because when taking the limit $e\to 1$ in Eq.~\ref{eq:A85}, we have $\tau_{\mathrm{A85}} \propto (e-1)^{3/2} \propto T_{f}$ as per Eq.~\ref{eq:cn}.
Similarly, using Eq.~\ref{eq:dn}, it is easy to see that $\lim_{e\to1} \tau_{\mathrm{A85}}$ is a constant at apocenter and thus a constant fraction of the orbital period.
We can also understand why in the bottom panel of Fig.~\ref{fig:timesteps}, the curve for $\tau_{\mathrm{A85}}$ is non-monotonic. 
This is because the expression uses $\y{4}$ and $\y{5}$ which have zero crossings for some~$e'$. 

With all of this in mind, we can now finally propose our new criterion, specifically we choose
\begin{eqnarray}
    \tau_{\mathrm{PRS23}} &\equiv& \sqrt 2 \;\cdot\;\left( \left(\frac{\y{3}}{\y{2}}\right)^2 + \left(\frac{\y{4}}{\y{2}}\right)  \right) ^{-1/2}\\
    &=&\sqrt {\frac{2\cdot\y{2}\y{2}}{\y{3}\y{3}+\y{2}\y{4}} }.
\end{eqnarray}
By construction, this expression avoids all the issues mentioned above.
We further choose the value of $\eta$ to be $\sqrt[7]{5040 \cdot \epsilon}$ so that we have
\begin{eqnarray}
dt_{\mathrm{PRS23}} &=& \sqrt[7]{5040 \cdot \epsilon} \; \cdot \; \tau_{\mathrm{PRS23}}. \label{eq:dtnew}
\end{eqnarray}
This results in identical timesteps for circular orbits using both the old and new criteria in IAS15. 
The scaling of $dt$ with $\epsilon$ is also the same using both new and old criteria.

Fig.~\ref{fig:timesteps} shows $dt_{\mathrm{PRS23}}$ as a function of the eccentricity (dashed black line). 
First, note that we indeed have $dt_{\mathrm{PRS23}} = dt_{\mathrm{RS15}}$ for circular orbits.
For high eccentricities $e\to 1$, the new criterion results in timesteps about a factor of~2.3~larger.
Second, note that in comparison to $dt_{\mathrm{RS15}}$ and $dt_{\mathrm{2,4}}$ there are no more singularities ($dt \to \infty$) for low eccentricity orbits near apocenter. 
This is because the term $\y{3}/\y{2}$ remains non-zero and thus dominates when $\y{4}/\y{2}$ is zero.
Third, note that in the limit of $e\to1$, the timestep at apocenter now converges towards a constant. This is because the term $\y{4}/\y{2}$ dominates in this limit.
Fourth, note that at pericenter the relevant timescale $T_f$ remains well resolved in the limit of $e\to1$.
This is because the term $\y{3}/\y{2}$ dominates in this limit.
Fifth, although $dt_{\mathrm{PRS23}}$ is not strictly monotonic as a function of eccentricity\footnote{A constant or at least a monotonic function of the eccentricity would be ideal here from a performance perspective because there is no physical timescale in the problem that has a local maxima. 
However, there are no free parameters to change in our current scheme. 
The only way to make further progress would be to include more derivatives which we can then assign different weights to or combine in other non-trivial ways. 
But using higher order derivatives comes at the disadvantage of making the scheme susceptible to floating point precision. 
For those reasons, we don't see an obvious way to improve the non-monotonicity.}, its local maxima at $e\approx 0.4$ is much less pronounced than in $dt_\mathrm{A85}$.

Fig.~\ref{fig:timestepshyper} is similar to Fig.~\ref{fig:timesteps} but shows the timesteps at pericenter for hyperbolic orbits as a function of periapse distance over semi-major axis. 
Orbits on the right are effectively a straight line, and those on the left make an almost $180^{\circ}$ turn.
Because hyperbolic orbits have no orbital timescale, we plot the timestep in units of the pericenter passage timescale, defined as $\y{1}/\y{0}$ at pericenter (and thus consistent with the pericenter passage timescale for circular orbits). 
As one can see, both the new and old timestepping criterion for IAS15 as well as all the other criteria always resolve the pericenter timescale in hyperbolic orbits well. 
As for the eccentric case, the new timestepping criterion results in timesteps about 2.3~times larger for the default value of $\epsilon$.
Of course, it is straightforward to construct a hyperbolic orbit that is not resolved correctly. 
This is not a shortcoming of a particular timestepping criterion but simply follows from the fact that there is no timescale other than the pericenter passage timescale in the problem\footnote{A setup that fails can be constructed by choosing the initial timestep of a hyperbolic encounter so large that the entire encounter occurs during one timestep. This is not possible for bound eccentric orbits because the orbital period puts an upper limit on the timestep.}. 
In practice, this is unlikely to be a problem because there are typically more than two particles being integrated at any given time.
The fact that every criterion can resolve hyperbolic encounters equally well is likely a reason why authors who are mostly interested in stellar systems without binaries sometimes do not find a significant difference when testing timestepping criteria \citep{Makino1991}.

One disadvantage of $\tau_{\mathrm{PRS23}}$ is that it diverges for $\y{2}=0$.
In practice this might not be an issue because in most implementations a timestep is only allowed to increase or decrease by some safety factor during one timestep. 
This safety factor is already mentioned in \cite{Aarseth1985} and is also used in IAS15. 
So even if we can't estimate a timescale because instantaneously $\y{2}=0$, the timestep will not blow up immediately.
Note that it is reasonable to assume that $\y{2}=0$ only instantaneously (e.g. at the end of one timestep).
If $\y{2}=0$ for more than one timestep, then the particle simply moves along a straight line for an extended period of time which implies that all the other derivatives are zero as well.
In that case the integrator will have no problem integrating the particle's motion accurately regardless of the timestep.
The only way to make a timestepping criterion well defined when $\y{2}=0$ and keep all the other desirable properties is to use higher derivatives, at least $\y{5}$ such as the criterion proposed by \cite{Aarseth1985}.
However, although $\tau_{\mathrm{A85}}$ is well defined if $\y{2}=0$, it is still not well defined if at some point $\y{2}=\y{3}=0$.
We are unable to construct a physical scenario in which $\y{2}=0$ but $\y{3}\neq 0$ for more than one timestep.
And as we have mentioned above and will show in tests below, having higher derivatives would make any criterion more susceptible to floating point precision issues. 
In summary, we have good reasons to think that $\tau_{\mathrm{PRS23}}$ is well behaved in almost all imaginable scenarios.

As a side note, the previous timestepping criterion in IAS15 calculated the new timestep somewhat inconsistently using a mix of accelerations from the beginning of the timestep and values from the series expansion in the middle of the timestep. 
With $\tau_{\mathrm{PRS23}}$, IAS15 is now relying on the internal series expansion to consistently estimate all derivatives at the end of the timestep.

\section{Tests}
\label{sec:tests}
The discussion above provided the theoretical motivation for our specific choice for $dt_{\mathrm{PRS23}}$.
In this section we present results from numerical tests, verifying that simulations with our new timestepping criterion are converged in a wide range of scenarios. 

\begin{figure}
    \centering
    \includegraphics[width=\columnwidth]{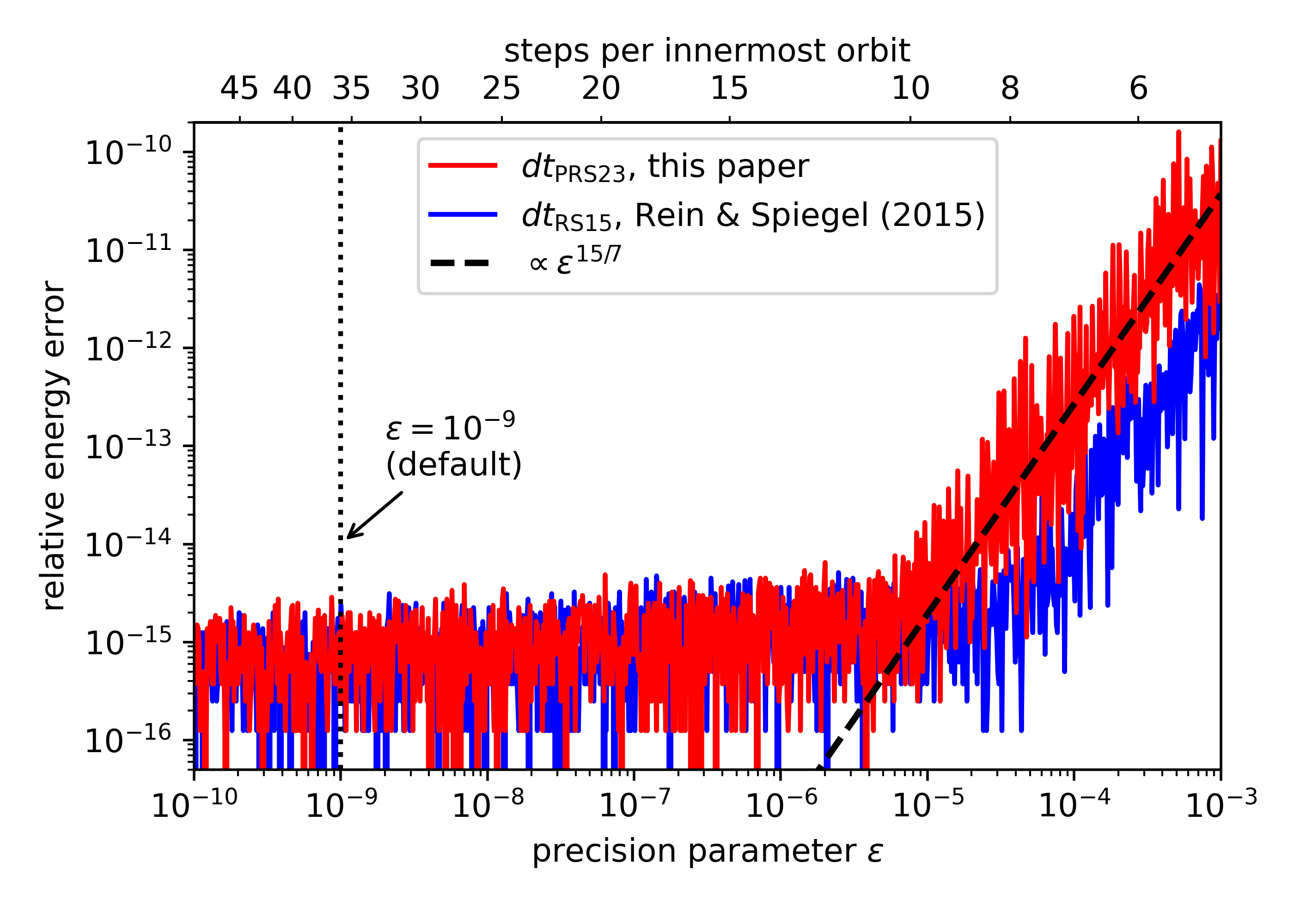}
    \caption{Relative energy error in simulations of the outer Solar System after 100 Jupiter orbits as a function of the accuracy control parameter $\epsilon$. The blue curve shows the error in simulations using timesteps given by $dt_{\mathrm{RS15}}$. The red curve shows the error in simulations using $dt_{\mathrm{PSR23}}$. Also shown is the expected power law scaling for a 15th order method.}
    \label{fig:epsilon}
\end{figure}

As a first test, we integrate the outer Solar System for 100 Jupiter orbits using different values of $\epsilon$ (and thus $\eta$ via Eq.~\ref{eq:dtnew}). 
The relative energy error is shown in Fig.~\ref{fig:epsilon} for both the old and new IAS15 timestepping criteria. 
One can see that for $\epsilon \lesssim 10^{-5}$, the results are converged to machine precision.
The default value of $\epsilon=10^{-9}$ is well within this regime.
Note that the relative error continues to decrease for $\epsilon<10^{-5}$.
This is because IAS15 uses compensated summation \citep{Kahan1965} for some internal values.
For $\epsilon \gtrsim 10^{-5}$, the error grows quickly.
Because IAS15 is a 15th order schema and the timestep is proportional to $\epsilon^{1/7}$, we have $\Delta E/E \propto \epsilon^{15/7}$.  
For the default $\epsilon=10^{-9}$, the timestep corresponds to about 35 timesteps per innermost orbital period in this problem.

\begin{figure}
    \centering
    \includegraphics[width=\columnwidth]{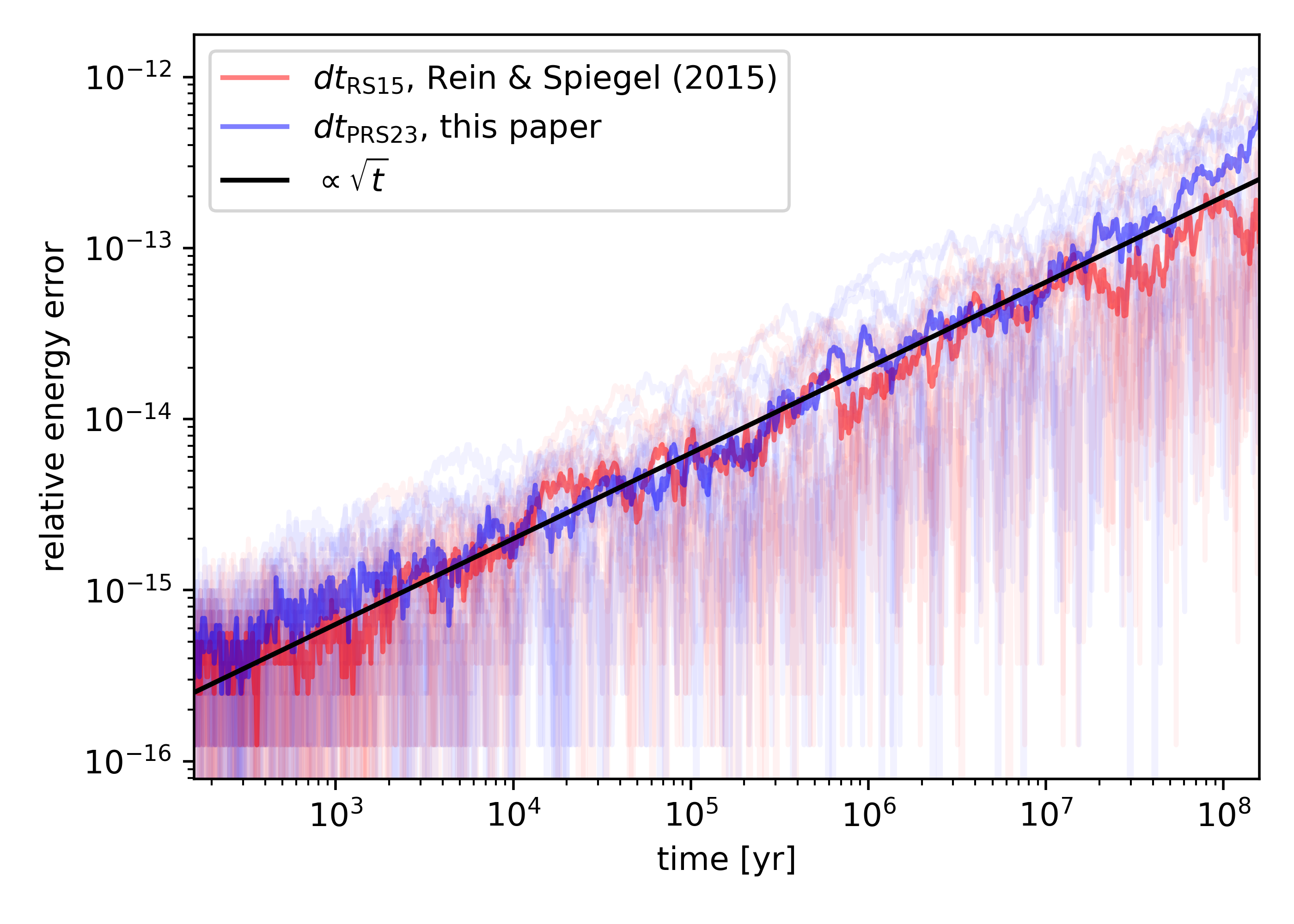}
    \caption{Relative energy error in 24 simulation of the outer Solar System as a function of time using the default $\epsilon=10^{-9}$. 
    The red curves show the error in simulations using timesteps given by $dt_{\mathrm{RS15}}$.
    The blue curves show the error in simulations using timesteps given by $dt_{\mathrm{PRS23}}$.
    Shown in black is the scaling expecting if the simulations follow Brouwer's law.
    }
    \label{fig:longterm}
\end{figure}
Next, we integrate the outer Solar System for $10^{8}$~years, or $10^{7}$~innermost periods.
We run 12 simulations using the old timestepping criterion, $dt_{\mathrm{RS15}}$, and 12 using the new criterion $dt_{\mathrm{PRS23}}$.
The relative energy errors of all individual simulations are shown as faint lines in Fig.~\ref{fig:longterm}. 
We also plot the median as a stronger line.
The results show that over these timescales, the energy error grows as the square root of time. 
This confirms that, at least under these conditions, IAS15 satisfies Brouwer's law \citep{Brouwer1937} regardless of the precise timestepping criterion.
Note that over much longer timescales, or by changing $\epsilon$, IAS15 will eventually show a linear error growth.
This is expected because IAS15 uses a series expansion internally.
This series has to be cut off after a finite number of terms so it cannot be unbiased for all times.
However, we don't expect this to matter in most astrophysical applications.

\begin{figure}
    \centering
    \includegraphics[width=\columnwidth]{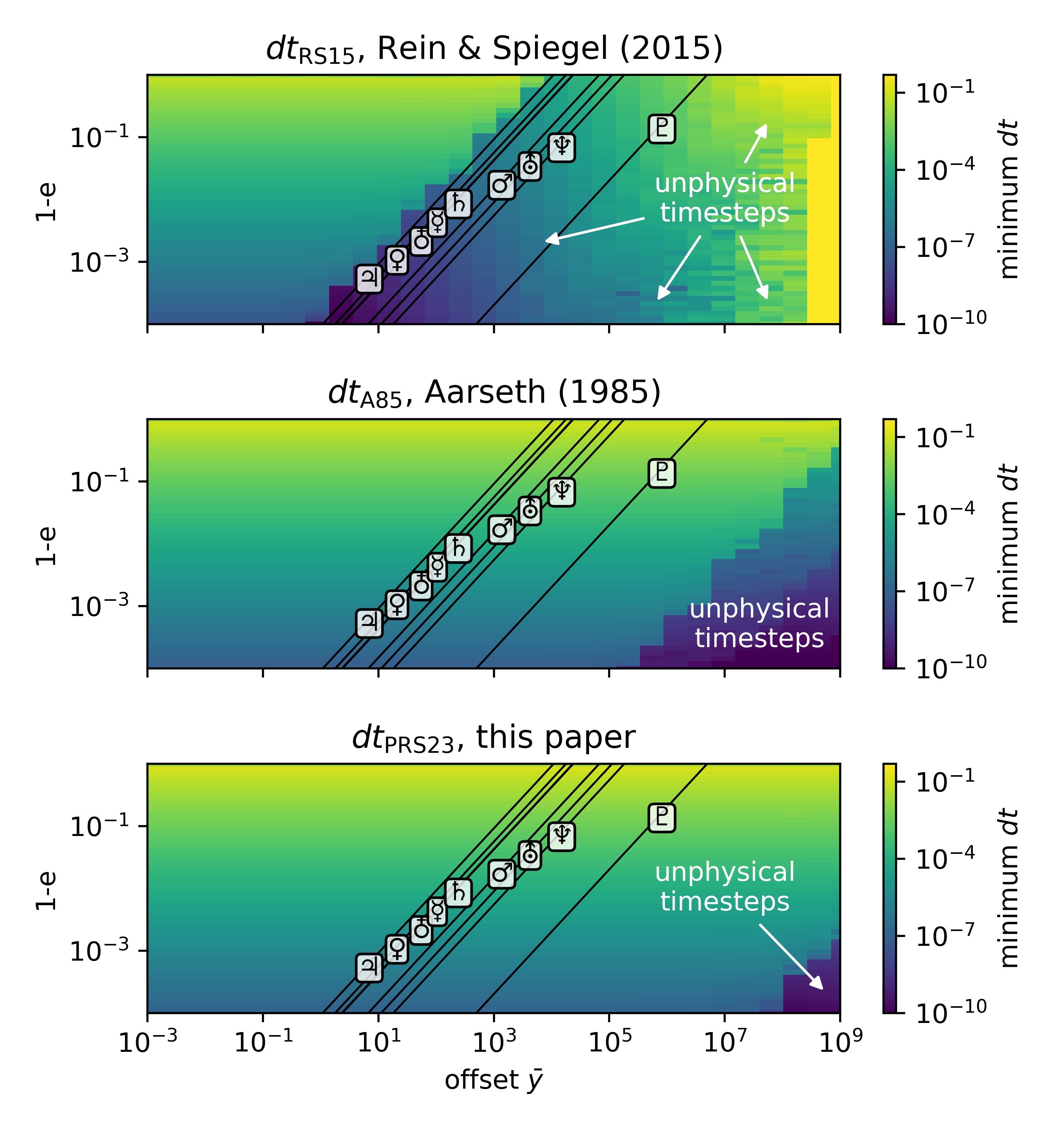}
    \caption{Minimal timesteps used in a two body problem as a function of eccentricity $e$ and offset from the coordinate origin~$\bar y$.
    We work in units where $G=a=1$.
    The top, middle, and bottom panels show the minimal timesteps in simulations using $dt_{\mathrm{RS15}}$, $dt_{\mathrm{A85}}$, and $dt_{\mathrm{PRS23}}$ respectively.
    The diagonal lines correspond to orbits around the Solar System planets whose pericenter touches the planet surface.
    Simulations using $dt_{\mathrm{PRS23}}$ are able to accurately calculate timesteps for offsets $10^8$~times larger than simulations using $dt_{\mathrm{RS15}}$.
    }
    \label{fig:offset}
\end{figure}

We now compare how well the new timestepping criterion, $dt_{\mathrm{PRS23}}$, can determine a physically meaningful timestep in an eccentric two-body system if the system is offset from the origin. 
As discussed above, this is a common scenario in N-body simulations, for example when planets orbiting a star have a close encounter with each other. 
For this test we initialize an eccentric two-body system on a grid with different offsets~$\bar y$ and eccentricities~$e$. 
We plot the minimum timesteps as given by $dt_{\mathrm{RS15}}$ in the top panel of Fig.~\ref{fig:offset} and those by $dt_{\mathrm{PRS23}}$ in the bottom panel.
Physically, the minimum timestep should only depend on the eccentricity, but not the offset. 
However, as one can see in Fig.~\ref{fig:offset}, $dt_{\mathrm{RS15}}$ quickly fails to determine correct timesteps. 
Even at moderate offsets of only 100 times the semi-major axis, $\bar y = 100\, a$, the algorithm fails at eccentricities higher than 0.99.
The reason for this is finite floating point precision as explained in Sect.~\ref{sec:floatingpoint}.
Because our new timestepping criterion, $dt_{\mathrm{PRS23}}$, uses lower derivatives the reliability is much improved.
We can now have offsets $10^{8}$ times larger before finite floating point precision becomes an issue.
The middle panel in the figure shows the $dt_{\mathrm{A85}}$ timestepping criterion. 
The criterion fails later than $dt_{\mathrm{RS15}}$ but earlier than $dt_{\mathrm{PRS23}}$. 
This is expected because $dt_{\mathrm{A85}}$ uses 5th order derivatives versus 7th and 4th order for the two other criteria respectively.

In Fig.~\ref{fig:offset} we also over-plot lines corresponding to particle orbits whose pericenter touches the planet's surface, assuming parameters for the planets in our Solar System. 
As one can see, the original $dt_{\mathrm{RS15}}$ criterion is not able to reliably integrate these orbits for any eccentricity and any Solar System planet which severely limits its usability. 
Both, $dt_{\mathrm{A85}}$ and $dt_{\mathrm{PRS23}}$ on the other hand can successfully determine the correct timescale.
We therefore expect them to be able to handle most simulations in the context of the Solar System and planetary dynamics in general.
Note that the problem gets worse for smaller objects further away from the central object (Pluto, planet 9, Kuiper belt objects).
For example, $dt_{\mathrm{A85}}$ can resolve orbits around a 10km sized particle located at Pluto's semi-major axis.
In contrast, our new criteria $dt_{\mathrm{PRS23}}$ can resolve orbits around particles at the same location but with a diameter as small as~50m.

Note that although the test above shows that we can resolve tight orbits around planets, the timestep chosen might not always be ideal.
Our timestep criterion normalizes~$\y{3}$ and~$\y{4}$ by comparing them to the acceleration~$\y{2}$.
There are cases where~$\y{2}$ is dominated by the gravitational force from the central object even though the particle is in a bound orbit around a planet. 
This might result in a timestep that is too large. 
As an example, consider the Earth-Moon-Sun system. 
The acceleration on the Moon from the Sun is twice as large as the acceleration from the Earth. 
As a result, the timestep in a simulation of the Earth-Moon-Sun system is about twice as large compared to a simulation of the Earth-Moon system alone. 
This is clearly not physical.
One can reduce $\epsilon$ if this becomes an issue. 
Another possible solution is to calculate $\tau_{\mathrm{PRS23}}$ for each particle pair in the simulation and then take the minimum. 
However, the latter increases the computational complexity.

\begin{figure}
    \centering
    \includegraphics[width=\columnwidth]{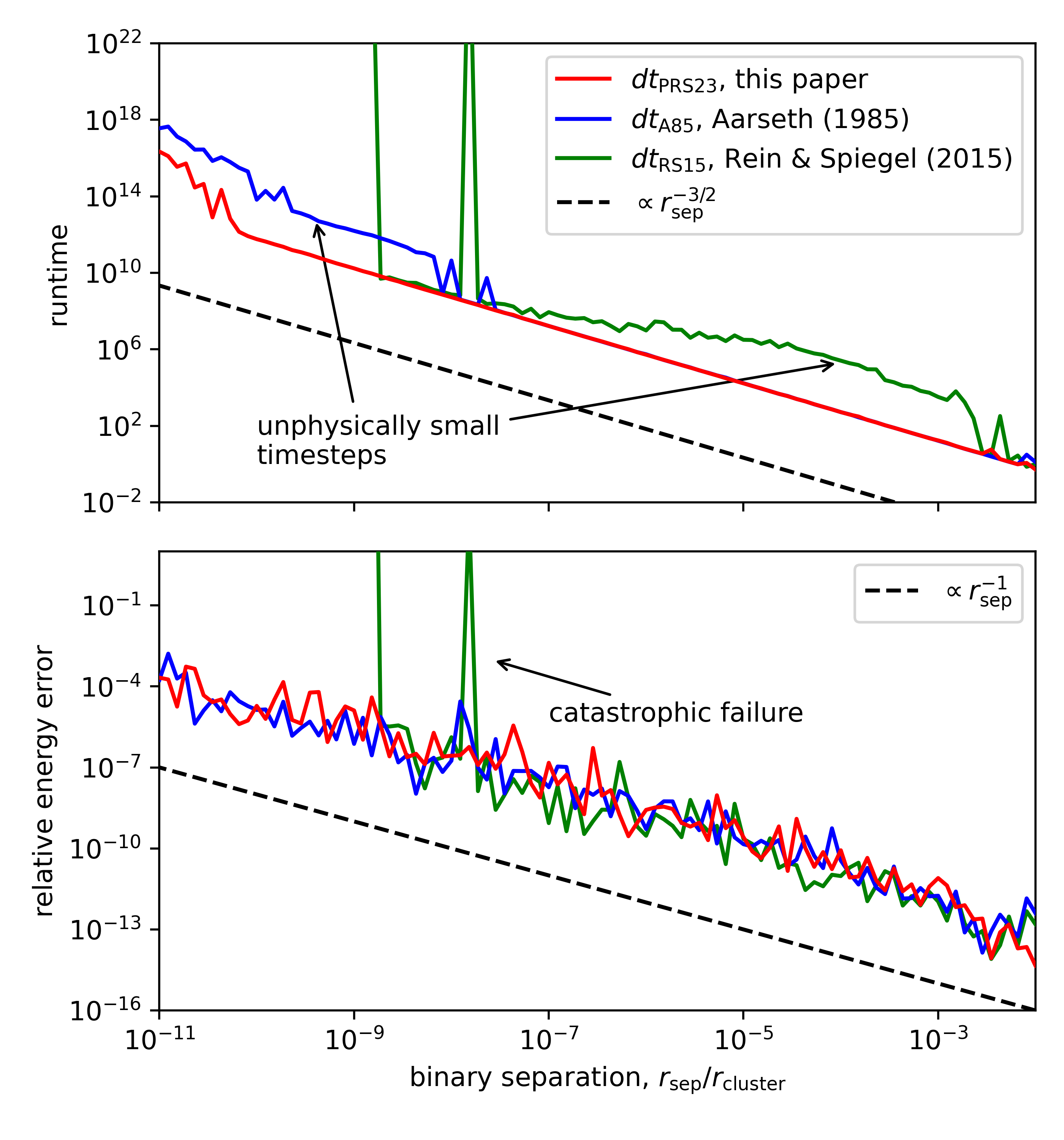}
    \caption{
    Runtime and relative energy error in a simulation of a stellar cluster with 50 binaries. 
    The red curves show simulations using our new $dt_{\mathrm{PRS23}}$ criterion which performs best in this test.
    }
    \label{fig:cluster}
\end{figure}

Whereas this paper focuses mostly on applications related to planetary systems, the timestepping criterion by \cite{Aarseth1985} was initially intended for N-body simulation of stellar clusters.
Thus, as a further test we integrate a stellar cluster. 
We initialize the cluster as a Plummer sphere with $N=50$ equal mass particles following the prescription of \cite{Aarseth1974}. 
To better illustrate the differences between the timestepping criteria, we then replace each particle with an equal mass circular binary (we thus have $N=100$ particles in total). 
We run 300 different realizations and vary the initial binary separation relative to the cluster size, $r_{\mathrm{sep}}/r_{\mathrm{cluster}}$.
The runtime (normalized such that it is 1 for $r_{\mathrm{sep}}/r_{\mathrm{cluster}}=10^{-2}$) and the relative energy errors for the timestepping criteria $dt_{\mathrm{RS15}}$, $dt_{\mathrm{A85}}$ and $dt_{\mathrm{PRS23}}$ are shown in Figure~\ref{fig:cluster} as a function of the binary separation.
The $dt_{\mathrm{RS15}}$ criterion performs worst. 
It chooses a timestep that is too pessimistic (small) by an order of magnitude even for wide binaries, $r_{\mathrm{sep}}/r_{\mathrm{cluster}} \lesssim 10^{-2}$, and then fails catastrophically at moderately tight binaries, $r_{\mathrm{sep}}/r_{\mathrm{cluster}} \lesssim 10^{-8}$.
As expected, our new criterion $dt_{\mathrm{PRS23}}$ performs significantly better, choosing the optimal timestep to resolve the binaries' orbital period down to $r_{\mathrm{sep}}/r_{\mathrm{cluster}} \lesssim 10^{-10}$, and never failing catastrophically in any of our tests involving stellar clusters.
The $dt_{\mathrm{A85}}$ criterion falls once again in-between the other two criteria.
Note that the relative energy error increases for decreasing binary separation due to finite floating point precision as $\propto r_{\mathrm{sep}}^{-1}$ regardless of the timestepping criteria as long as the criterion resolves the orbital period of the binaries.

We attribute the results in our cluster tests to the same issues discussed in detail above for the three body systems.
Most importantly, we find that tight binaries and close encounters in stellar cluster can pose difficulties for some timestepping criteria, resulting in unphysically small timesteps and thus causing simulations to slow down significantly.
Once floating point errors dominate, simulations can also fail catastrophically. 
This is why the criteria with the lowest order derivative, $dt_{\mathrm{PRS23}}$, which is least susceptible to these errors, performs best.

\section{Conclusions}
\label{sec:conclusions}
In this paper, we have described a new timestepping criterion and implemented it for the IAS15 integrator in the REBOUND N-body package. 
Compared to the previous timestepping criterion in IAS15, our new criterion has several significant advantages.

First, we provide a physical justification for our choice of timestep. 
Specifically, we choose the timestep so that for any bound orbit it is a fraction of the orbital period at apocenter and a fraction of the pericenter timescale at pericenter. 
In the limit of large eccentricities, these fractions are well behaved and converge to a constant, independent of eccentricity. 
We have shown that our new criterion is the simplest choice (using the lowest order derivatives) that can satisfy these requirements.
    
Second, by using low (second, third, and fourth) order derivatives of the coordinates, we are able to avoid most issues coming from finite floating point precision. 
This allows us to accurately choose timesteps for particles having close encounters $10^8$ times further away from the coordinate origin compared with the original timestepping algorithm in IAS15.
Specifically, we can resolve orbits around a particle with a 50m diameter located at Pluto's current semi-major axis.
We expect that this will be the most noticeable change for users that previously had trouble resolving tight orbits, close encounters, Zeipel-Kozai-Lidov cycles \citep{Zeipel1910,Kozai1962,Lidov1962}, or other scenarios where high eccentricities occur.

Third, although our criterion is similar to that of \cite{Aarseth1985}, it is simpler in the sense that we do not use the 5th derivative and it is therefore less susceptible to floating point precision issues.

Finally, let us make two important remarks about adaptive step size control in N-body simulations.
First, it is always possible to come up with a scenario where an adaptive timestepping criterion fails. 
For example, if the initial timestep is orders of magnitudes too large, or timescales in a simulations change very abruptly, then any criterion will fail.
So some physical intuition is always required when setting up and analysing N-body simulations with adaptive timestepping.

Second, it can be tempting to try and speed up simulations by changing the precision parameter in an adaptive scheme.
For IAS15 this would involve adjusting $\epsilon$. 
However, note that increasing $\epsilon$ by a factor of 10 only increases the timestep by a factor of~$1.4$. 
The energy error on the other hand increases by a factor of~140, a consequence of IAS15 being a 15th order scheme (see right side of Fig.~\ref{fig:epsilon}).
Thus, one very quickly moves the integrator out of the regime where it follows Brouwer's law. 
As soon as this happens, the energy error will grow linearly in time, rendering some long term integrations unreliable.
In summary, one should take great care and perform a convergence study when experimenting with larger than default timesteps in IAS15 and other high order integrators. 
In most cases the speedup of a factor of a few ($\lesssim 5$) might not be worth the extra effort.

Compared to the existing timestepping criterion in IAS15, our new criterion is better in all cases, and it will therefore become the default for new simulations.  
Users who wish to continue to use the old criterion can set the new \texttt{ri\_ias15.adaptive\_mode} flag to 1. 
We also implement the \cite{Aarseth1985} criterion in IAS15 which can be turned on by setting \texttt{ri\_ias15.adaptive\_mode=3}.
REBOUND and IAS15 are freely available at \url{https://github.com/hannorein/rebound}.

\section*{Acknowledgements}
This research has been supported by the Natural Sciences and Engineering Research Council (NSERC) Discovery Grant RGPIN-2020-04513.

\bibliography{references}

\onecolumngrid \footnotesize

\end{document}